# Magneto-capacitance probing of the many-particle states in InAs dots

Oliver S. Wibbelhoff[1], Axel Lorke[1], Dirk Reuter[2], and Andreas D. Wieck[2]

[1]Institute of Physics, Universität Duisburg-Essen, Lotharstr. 1, D-47048 Duisburg, Germany
[2]Solid State Physics, Ruhr-Universität Bochum, Universitätsstr. 150, D-44780 Bochum, Germany

We use frequency-dependent capacitance-voltage spectroscopy to measure the tunneling probability into self-assembled InAs quantum dots. Using an in-plane magnetic field of variable strength and orientation, we are able to obtain information on the quasi-particle wave functions in momentum space for 1 to 6 electrons per dot. For the lowest two energy states, we find a good agreement with Gaussian functions for a harmonic potential. The high-energy orbitals exhibit signatures of anisotropic confinement and correlation effects.

Tunneling spectroscopy is a well-known tool for the determination of quantized energies. Recently, it has also become possible to use this technique to directly study the wave function associated with an energy level. Using scanning tunneling microscopy (STM), it has been possible to observe the charge densities in structures like the quantum corral[1], image standing wave patterns with cross-sectional STM[2] or to map out the probability densities in freestanding InAs dots[3]. A complementary technique, which probes the probability density in $\underline{k}$-space, makes use of magneto-tunneling spectroscopy[4-5].

Self-assembled quantum dots (QDs) have been thoroughly investigated in the last few years, using electronic and optoelectronic methods[6-7]. The high interest in these nanoscopic systems stems from their potential application in laser devices or for quantum computing due to their controllable and discrete spectrum of energy levels.

Here we present a novel technique to map out the quasi-particle wave function for interacting electrons in self-assembled QDs, embedded in a field-effect transistor (FET) structure[8-9]. High-resolution maps of the wave functions are obtained using capacitance-voltage (CV) spectroscopy[7-9] in variable magnetic fields.

The samples contain InAs QDs embedded in a capacitor-like heterostructure between a Si-doped GaAs back contact and a surface gate. We have investigated two types of samples, which differ slightly in the thickness $d_1$ of the triangular tunneling barrier (see Fig. 1 (a)) that separates the dots from the back contact. The characteristic decay time of tunneling electrons[10] depends roughly exponentially on the thickness of the barrier, which is equal to $d_1 = 40$ nm for sample type A and $d_1 = 42.5$ nm for type B. For details of the layer structure and growth procedure, see e.g. [9].

For the CV spectroscopy, we apply a DC gate bias with an additional AC voltage (5 mV rms) of variable frequency $f$ and measure the capacitive response signal. Fig. 1(a) shows the conduction band profile of our heterostructure for two different DC voltages. With increasing bias, electronic dot states in the quantum well can be shifted with regard to the chemical potential $\mu_{bc}$ in the back contact so that, in resonance, tunneling of electrons back and forth between the reservoir and the dots occurs, leading to an increase in the measured capacitance. At sufficiently high frequencies, the AC modulation period will become shorter than the charging time, so that the capacitance maximum will be suppressed. It can be shown that the amplitude of the charging peak directly reflects the tunneling probability[10].

For the wave function mapping, we follow the approach of Patanè et al.[5] and apply a magnetic field $B$ perpendicular to the tunneling direction $z$ (see Fig. 1 (b)), at various angles in the $x$-$y$-plane.

Due to the Lorentz force caused by the magnetic field, tunneling electrons from the back contact into the dots will acquire an additional momentum[11]

$$\Delta k = \frac{d_1 eB}{\hbar}. \qquad (1)$$

Thus, the overlap of wave functions in the electron reservoir ($\varphi_{bc}(k-\Delta k)$) and the dots ($\varphi_{dot}(k)$) can be altered by changing the in-plane field $B$. As the capacitive signal is proportional to the tunneling probability, which in turn is given by the overlap of the wave functions, we are able to obtain information on the shape of $\varphi_{dot}(k)$.

Using capacitance spectroscopy rather than measuring the current through resonant tunneling diodes[5] has a number of advantages. First, we are measuring a large number of dots simultaneously, which gives us information on characteristic dot properties. Also, the large number of dots probed provides for a high signal-to-noise ratio[12].

Furthermore, the AC frequency $f$ can be adjusted to account for the fact that the tunneling current increases exponentially with increasing energy. Therefore, we can set $f$ to optimize the wave function mapping. Finally, in our FET structures, it is possible to control the number of electrons $N_e$ in the dots by simply changing the DC gate voltage. Due to superimposed AC voltage, only the last electron tunnels in and out of the dots. This way, we probe the quasi-particle wave function[13] of the last electron added to the dot, which gives information on interactions that govern the *many-particle* states in self-assembled dots.

By varying the magnetic field amplitudes between 0 and 11 T and the field direction within the $x$-$y$-plane in 20 steps of 18°, we obtain three-dimensional plots of the tunneling probability in $\underline{k}$-space.

Fig. 1(c) shows a series of CV traces for sample B with a tunneling barrier of $d_1 = 42.5$ nm at 4.2 K. The uppermost curves are typical low frequency traces[9] ($f = 35$ Hz), showing two distinct peaks ($s_1$, $s_2$) related to the filling of the so-called s-shell of the QDs[6-10,14], followed by four peaks[14] of the p-shell ($p_1 p_2 p_3 p_4$), which are not resolved here but coalesce into a broad shoulder. The lower CV traces (offset for clarity) are taken on the same sample at $f = 12$ kHz for various magnetic fields.

At sufficiently high frequencies, all charging peaks can be reduced in amplitude by application of a magnetic field, with a full suppression to the background capacitance at 8 T (lowest curve). Care has to be taken when choosing an ap-propriate modulation frequency. For each charging peak we choose a frequency which is high enough to ensure that the peak amplitude is given by the tunneling probability (rather than by the geometric parameters[10]) and at the same time low enough so that a good signal-to-noise ratio is obtained. We find that is this regime, the measured (normalized) probabilities are frequency independent.

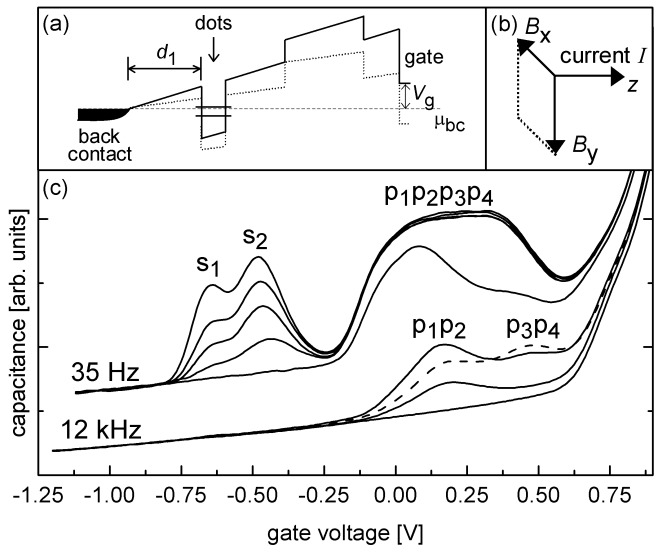

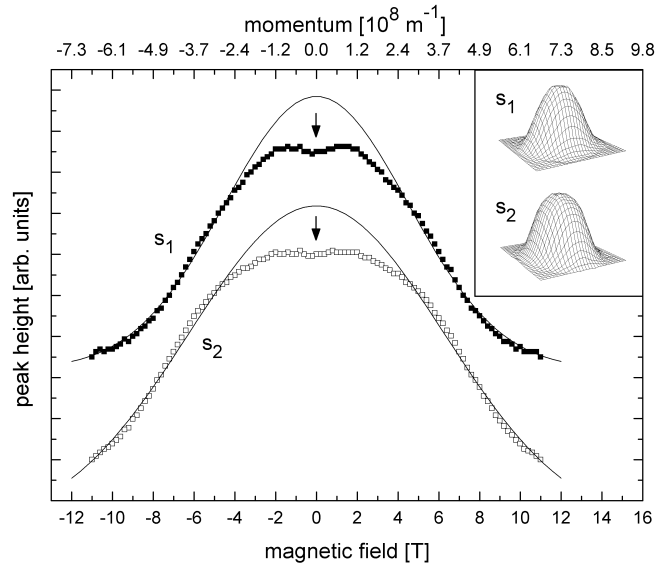

FIG. 1. (a) Conduction band profile of the heterostructure for two different biases, (b) directions of the current and magnetic field, where $x$-$y$ is the sample plane. (c) Low-temperature ($T = 4.2$ K) CV traces for sample B ($d_1 = 42.5$ nm) taken at low and high frequency. The curves show the influence of an in-plane magnetic field on the s-states (0 T, 3 T, 4 T, 5 T and 8 T, top to bottom) and on the p-states (0 T (dashed), 2 T, 5 T and 8 T).

FIG. 2. Tunneling probabilities for the first ($s_1$) and second ($s_2$) electron in InAs quantum dots (sample A) in $\underline{k}$-space, where $s_2$ is offset for clarity. Filled squares indicate measured data for $s_1$ and empty squares data for $s_2$, the lines are Gaussian fits. The corresponding CV data was measured at 5.1 kHz. The inset shows the full rotational plots of the measured probability densities.

To map out the s-states, we have plotted in Fig. 2 the CV peak heights for $s_1$ and $s_2$ (sample A) as a function of the magnetic field. The top scale in the figure gives the corresponding momentum evaluated from Eq. (1). At the edges of the plots, the data agree quite well with the commonly used parabolic model[5-7,13-15], however, we observe characteristic deviations at the center of the wave function (arrows). For small magnetic fields, tunneling can actually be increased, resulting in a center dip in the probability densities.

A comparison between both s-states is of interest as it directly shows the influence of the additional electron on the measured probability. Because of the repulsive electron-electron interaction, one would intuitively expect that the wave function should occupy a larger volume in real space when a second electron is added to the dot. Similarly, due to the higher energy for the doubly occupied $s_2$ state, the corresponding wave function is expected to spread further into the confining potential. We observe, however, that the state $s_2$ (FWHM: $7.6 \cdot 10^8$ m$^{-1}$) is more extended in $\underline{k}$-space, and therefore smaller in real space than the $s_1$ state (FWHM: $5.5 \cdot 10^8$ m$^{-1}$). One possible explanation could be the fact that the $s_2$ state corresponds to a filled electronic shell and that – similarly as in atomic physics – configurations with completely filled shells have smaller electronic radii than singly occupied orbitals. However, calculations using a *parabolic* confinement[13] do not reproduce this behavior and a more detailed investigation is needed to clarify its origin.

The charging peaks of the p-states are not as clearly separated as those of the s-states and can therefore not be mapped individually. We find, however, when a magnetic field is applied, that the $p_1$ and $p_2$ peaks show a distinctly different behavior from the $p_3$ and $p_4$ peaks (cf. Figure 1(c)). This reflects the fact that here, the electrons corresponding to $p_1$ and $p_2$ ($p_3$ and $p_4$) are in the same orbital state, respectively. We therefore evaluate the two lower ($p_1p_2$) and two higher ($p_3p_4$) states jointly in the following.

Fig. 3 shows maps of the respective tunneling probabilities (sample B). In both cases, as expected for the p-states in $x$-$y$-representation, we observe a non-monotonic behavior with a minimum in the center and two maxima, which are in orthogonal directions for the two different pairs. However, unlike the single particle states in the harmonic oscillator model, the minima in the center do not fall to zero but still indicate finite tunneling probabilities for $B = 0$.

A comparison of the plots for $p_1p_2$ and $p_3p_4$ shows, in analogy to the s-states, a wave function that extends further in $\underline{k}$-space for $N_e = 5 - 6$ (completely filled p-shell) than for $N_e = 3 - 4$.

The consecutive filling of the same spatial orbital for the two sequences $p_1p_2$ and $p_3p_4$, respectively, shows in a surprisingly clear way that the commonly assumed circular or square symmetry of the dots is lifted so that Hund's rule[16] can not be applied. In the contour plots in Fig. 3, this is demonstrated by the oval shape of the probability densities, clearly oriented along the $\langle 110 \rangle$ crystal axes. The lower states ($p_1p_2$) are always elongated along the [011] direction in real space. The lifting of degeneracy can either be attributed to piezoelectric effects[7,17-18] or to a slight elongation of the InAs island shape[15]. A piezoelectric potential is induced in the material by strain, which remains despite the relaxation processes that take place during the growth procedure.

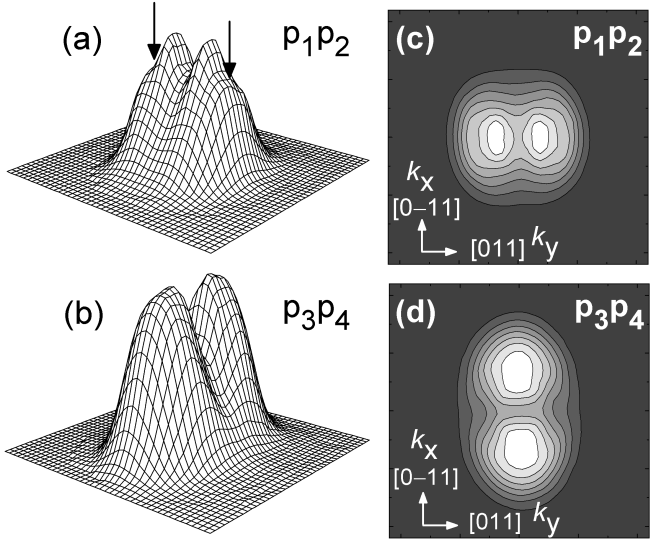

FIG. 3. (a) and (b) Surface plots of the quasi-particle wave function for sample B probed at a frequency of 12 kHz for the low-energy ($p_1p_2$) and high-energy p-states ($p_3p_4$). (c) and (d) Corresponding contour plots.

An elongation of the dot shape caused by anisotropic In diffusion during growth has been proposed before[15,19], but seems to be too small to be the only reason responsible for the noticeable effect here. Therefore, we expect an additional contribution from the piezoelectric potential, which has an anisotropic modulation[17] and breaks the symmetry in the dot.

Also, due to the high resolution of our plots, we can identify a fine structure in both maps, which is more pronounced in the $p_1p_2$-states. In addition to the sharp maximum along the [011] direction, we observe a shoulder at higher $k_y$-values (see arrows in Fig. 3(a)). This is a clear evidence for the complex electronic structure of the quasi-particle states.

In conclusion, we have demonstrated a new capacitive technique to map out the wave functions in InAs quantum dots. As the occupation number in the dots can be controlled, our approach gives access to the addition spectra of interacting electrons into many-particle states. We find that the quasi-particle wave function decreases in size when electronic shells are completed, in agreement with smaller atomic radii for filled shells in atomic physics.

Furthermore, we find a lifting of the degeneracy along the ⟨110⟩ crystal axes and a fine structure in the magneto-tunneling maps, which can so far not be explained.

The authors would like to thank Alexander Govorov, Peter Entel, Dietrich Wolf, Elisa Molinari and Massimo Rontani for valuable discussions and the latter also for making the data of Ref. [13] available to us. Financial support by the Deutsche Forschungsgemeinschaft is gratefully acknowledged.